\documentclass{PoS}

\title{Bulk thermodynamics and charge fluctuations at non-vanishing baryon density\\[-4cm]
\normalsize \tt \hspace*{11.3cm}BNL-NT-07/43\\
\normalsize \tt \hspace*{11.3cm}BI-TP2007/28\\[3cm]
}

\ShortTitle{Thermodynamics and fluctuations at non-vanishing density}

\author{\speaker{Chuan Miao and Christian Schmidt} (RBC-Bielefeld Collaboration) \\
        Falkut\"at f\"ur Physik, Universit\"at Bielefeld, D-33615 Bielefeld, Germany\\
        Brookhaven National Laboratory, Upton, NY 11973, USA\\
        E-mail: \email{chuan@physik.uni-bielefeld.de, cschmidt@quark.phy.bnl.gov}}

      \abstract{We present results on bulk thermodynamic quantities as well as
        net baryon number, strangeness and electric charge fluctuations in QCD
        at non-zero density and temperature obtained from lattice calculations
        with almost physical quark masses for two values of the lattice cut-off
        $aT=1/4$ and $1/6$ . We show that with our improved p4fa3-action the
        cut-off effects are under control when using lattices with a temporal
        extent of 6 or larger and that the contribution to the equation of
        state, which is due to a finite chemical potential is small for
        $\mu_q/T<1$. Moreover, at vanishing chemical potential, i.e.  under
        conditions almost realized at RHIC and the LHC, quartic fluctuations of
        net baryon number and strangeness are large in a narrow temperature
        interval characterizing the transition region from the low to high
        temperature phase. At non-zero baryon number density, strangeness
        fluctuations are enhanced and correlated to fluctuations of the net
        baryon number. If strangeness is furthermore forced to vanish, as it
        may be the case in systems created in heavy ion collisions, strangeness
        fluctuations are significantly smaller than baryon number
        fluctuations.}

\FullConference{The XXV International Symposium on Lattice Field Theory\\
                 July 30-4 August 2007\\
                 Regensburg, Germany}
\usepackage{amstext}
\usepackage{graphicx}
\usepackage{pifont}
\usepackage{amsmath}

\begin{document}

\section{Introduction}
Heavy-ion collision experiments at RHIC and LHC lead to thermalized dense
matter at small but non-zero baryon density, or equivalently chemical
potential. Therefore it is necessary to study the bulk thermodynamics of QCD at
finite chemical potentials. In this work, we use the Taylor expansion method
\cite{method} to study the equation of state, the number density and
fluctuations of various quantum numbers on the lattice. We study \(2+1\) flavor
QCD with tree level Symanzik-improved gauge action and p4fat3-improved
staggered fermion action \cite{Karsch:2000kv}.  The simulations are carried out
on \(16^{3}\times4\) and \(24^{3}\times6\) lattices on a line of constant
physics with almost physical quark masses; the pion mass is about \(220\) MeV
and the strange quark mass is adjusted to its physical value.  We have scanned
a temperature range approximately from \(170\) MeV to \(500\) MeV. We are using
the exact RHMC algorithm \cite{RHMC} to update configurations. Details on our
simulation parameters can be found in \cite{Jan}.

\section{Taylor expansions of thermodynamic quantities}
For a large homogeneous system, the pressure of  QCD with \(u\), \(d\) and
\(s\) quarks can be expressed
as\begin{equation}
\frac{p}{T^{4}}=\frac{1}{VT^{3}}\ln Z\left(V,T,\mu_{u},\mu_{d},\mu_{s}\right),
\label{eq:Pdef}
\end{equation}
where the partition function $Z$ is a function of the volume $V$, temperature
$T$ and chemical potentials of $u$, $d$, and $s$ quarks. We have not considered
other species of quarks whose masses are much heavier. Due to the sign problem,
the difficulty of a direct lattice calculation at non zero chemical potentials
arises. We perform a Taylor expansion in terms of the chemical potentials
\begin{equation}
\frac{p}{T^{4}}
=\sum_{i,j,k}c_{ijk}(T)\left(\frac{\mu_{u}}{T}\right)^{i}
\left(\frac{\mu_{d}}{T}\right)^{j}\left(\frac{\mu_{s}}{T}\right)^{k},
\label{eq:PTaylor}
\end{equation}
and compute the coefficients $c_{ijk}$ at zero chemical potentials. When the
sum $i+j+k$ is odd, the coefficient $c_{ijk}$ is given as expectation value of
purely imaginary operators and therefore vanishes exactly. This reflects the
invariance of the QCD partition function under change of particle and
anti-particle. The leading term $c_{000}$ gives the pressure at vanishing
baryon density and can be calculated via the integral method. Results for the
parameter values considered here have been presented in \cite{Jan}. In
this work, we will concentrate on the part of the pressure
\begin{equation} 
\Delta p=p(\vec\mu)-p(\vec\mu=0),
\end{equation}
that arises due to non-zero chemical potentials, where
\(\vec\mu=(\mu_{u},\mu_{d},\mu_{s})\). For \(i+j+k>0\), the coefficients
\begin{equation}
c_{ijk}=\left.\frac{1}{i!j!k!}
\frac{\partial^{i}}{\partial\hat\mu_{u}^{i}}
\frac{\partial^{j}}{\partial\hat\mu_{d}^{j}}
\frac{\partial^{k}}{\partial\hat\mu_{s}^{k}}
\left( p/T^{4}\right)\right|_{\mu=0},
\end{equation}
are derivatives of the partition function, and can be calculated on the
lattice, where \(\hat\mu=\mu/T\).  These coefficients provide information about
other thermal quantities as well. For example, the strange quark number density
expands in chemical potentials as
\begin{equation}
\frac{n_{s}}{T^{3}}=\sum_{i,j,k} \left( k+1 \right) 
c_{ij(k+1)}\hat\mu_{u}^{i}\hat\mu_{d}^{j}\hat\mu_{s}^{k},
\label{eq:ns}
\end{equation}
and similarly the light up and down quark numbers. We can further consider
fluctuations in these quantities.

Alternatively, one can introduce chemical potentials for the conserved quantities
baryon number \(B\), electric charge \(Q\) and strangeness \(S\), which are
related to \(\mu_u, \mu_d ,\mu_s \) via
\begin{equation}
\mu_u=\frac{1}{3}\mu_{B} +\frac{2}{3}
\mu_{Q},\qquad
\mu_{d}=\frac{1}{3}\mu_{B}-\frac{1}{3}\mu_{Q},\qquad
\mu_{s}=\frac{1}{3}\mu_{B}-\frac{1}{3}\mu_{Q}-\mu_{S},
\label{eq:chempot}
\end{equation}
and compute e.g. the baryon density as 
\begin{equation} 
n_{B}=\frac{1}{3}\left( n_{u}+n_{d}+n_{s}\right).
\end{equation}
Then we can study densities and fluctuations in \(B \), \(Q\) and \(S\).

In the following, we will regard $u$ and $d$ quarks as degenerate and consider
$2+1$ flavor QCD. With the definition $\mu_{q}\equiv\mu_{u}=\mu_{d}$ for the
light quarks, the coefficients are
\begin{equation}
c_{ij}^{qs}=\left.\frac{1}{i!j!}
\frac{\partial^{i}}{\partial\hat\mu_{q}^{i}}
\frac{\partial^{j}}{\partial\hat\mu_{s}^{j}}
\left(p/T^{4}\right)\right|_{\vec \mu=\mathbf{0}},
\label{eq:cqs}
\end{equation} 
where the subscripts denote the order of the derivative and the superscripts
indicate the corresponding flavors.  If not specified, the default superscripts
will be \(qs\) and will often be left out. It is evident from Eqs.
(\ref{eq:chempot}) that choosing $\mu_u\equiv\mu_d$ is equivalent to a
vanishing electric charge potential $\mu_Q\equiv 0$.

Now we discuss how to evaluate these coefficients on the lattice. Inserting
Eq. (\ref{eq:Pdef}) into Eq. (\ref{eq:cqs}) and integrating out the fermion
fields in the partition function yields the coefficients as expectation
values of operators that contain derivatives of the determinant of the fermion
matrix \(M\). For example  the formula for \(c_{20}\) reads \begin{equation}
c_{20}
=\frac{N_{\tau}}{2N_{\sigma}^{3}}\left({\frac{n_{f}}{4}}
\left\langle {\frac{\partial^{2}\ln\det M}{\partial\vec \mu_{q}^{2}}}\right\rangle 
+{\left(\frac{n_{f}}{4}\right)^{2}}
\left\langle \left({\frac{\partial\ln\det M}{\partial\vec \mu_{q}}}\right)^{2}\right\rangle \right),
\end{equation}
where \(N_{\tau }\) and \(N_{\sigma}\) are temporal and spacial extent of the
lattice, \(n_{f} \) is the number of quark flavors in question (here
\(n_{f}=2\)), and \(\left\langle \cdots\right \rangle\) indicates taking the
thermal average over the ensemble. On each configuration, derivatives of
$\ln\det M$ need to be evaluated up to the same order as the order of the
expansion coefficients. These derivatives lead to the appearances of the
inverse fermion matrix \(M^{-1}\) inside traces
\begin{eqnarray}
\frac{\partial\ln\det M}{\partial\mu} 
& = & \mbox{Tr}\left({M^{-1}}\frac{\partial M}{\partial\mu}\right),\\
\frac{\partial^{2}\ln\det M}{\partial\mu^{2}} 
& = & \mbox{Tr}\left({M^{-1}}\frac{\partial^{2}M}{\partial\mu^{2}}\right)
-\mbox{Tr}\left({M^{-1}}\frac{\partial M}{\partial\mu}{M^{-1}}\frac{\partial M}{\partial\mu}\right).
\end{eqnarray}
To avoid full matrix inversions, we use the random noise method in estimating
such traces. Suppose we have generated a set of \(N\) random noise vectors \(R^{(a)},
a=1,\ldots,N\), then the trace can be estimated as
\begin{equation}
\text{Tr}\left( \mathcal{O}\ M^{^{-1}}\right)
\approx\frac{1}{N}\sum^{N}_{a=1}R^{(a)}\mathcal{O}M^{-1}R^{(a)},
\end{equation}
where \(\mathcal{O}\) is some arbitrary matrix. For each vector \(R^{(a)}\)
only the linear system \(MX=R^{(a)}\) needs to be solved. It is still quite
expensive to compute all necessary operators, since a large number of
random vectors is needed in order to get a satisfactory accuracy. Also, higher order
coefficients are more expensive, because more operators are needed. For the 4th
order coefficients one has
\begin{equation}
\begin{split}
c_{40} \ = \ &\frac{1}{4!N_{\sigma}^{3}N_{\tau}}\left\{ {\frac{n_{f}}{4}}\left\langle {\frac{\partial^{4}\ln\det M}{\partial\mu_{q}^{4}}}\right\rangle\right.\\
&  +4{\left(\frac{n_{f}}{4}\right)^{2}}\left\langle {\frac{\partial^{3}\ln\det M}{\partial\mu^{3}_{q}}}\frac{\partial\ln\det M}{\partial\mu_{q}}\right\rangle+3{\left(\frac{n_{f}}{4}\right)^{2}}\left\langle \left(\frac{\partial^{2}\ln\det M}{\partial\mu^{2}_{q}}\right)^{2}\right\rangle\\ &+6{\left(\frac{n_{f}}{4}\right)^{3}}\left\langle \frac{\partial^{2}\ln\det M}{\partial\mu^{2}_{q}}\left(\frac{\partial\ln\det M}{\partial\mu_{q}}\right)^{2}\right\rangle +{\left(\frac{n_{f}}{4}\right)^{4}}\left\langle \left(\frac{\partial\ln\det M}{\partial\mu_{q}}\right)^{4}\right\rangle \\
   &\left. -3\left({\frac{n_{f}}{4}}\left\langle \frac{\partial^{2}\ln\det M}{\partial\mu^{2}_{q}}\right\rangle +{\left(\frac{n_{f}}{4}\right)^{2}}\left\langle \left(\frac{\partial\ln\det M}{\partial\mu_{q}}\right)^{2}\right\rangle \right)^{2}\right\}, 
\end{split}
\end{equation} 
where 
\begin{eqnarray}
\frac{\partial^{3}\ln\det M}{\partial\mu^{3}} & = & \mbox{Tr}\left(M^{-1}\frac{\partial^{3}M}{\partial\mu^{3}}\right)-3\mbox{Tr}\left(M^{-1}\frac{\partial M}{\partial\mu}M^{-1}\frac{\partial^{2}M}{\partial\mu^{2}}\right)\notag\\
&&+2\mbox{Tr}\left(M^{-1}\frac{\partial M}{\partial\mu}M^{-1}\frac{\partial M}{\partial\mu}M^{-1}\frac{\partial M}{\partial\mu}\right),\\
\frac{\partial^{4}\ln\det M}{\partial\mu^{4}} & = & \mbox{Tr}\left(M^{-1}\frac{\partial^{4}M}{\partial\mu^{4}}\right)-4\mbox{Tr}\left(M^{-1}\frac{\partial M}{\partial\mu}M^{-1}\frac{\partial^{3}M}{\partial\mu^{3}}\right)\notag\\
 &  & \mspace{-36.0mu}-3\mbox{Tr}\left(M^{-1}\frac{\partial^{2}M}{\partial\mu^{2}}M^{-1}\frac{\partial^{2}M}{\partial\mu^{2}}\right)+12\mbox{Tr}\left(M^{-1}\frac{\partial M}{\partial\mu}M^{-1}\frac{\partial M}{\partial\mu}M^{-1}\frac{\partial^{2}M}{\partial\mu^{2}}\right)\notag\\
 &  & \mspace{-36.0mu}-6\mbox{Tr}\left(M^{-1}\frac{\partial M}{\partial\mu}M^{-1}\frac{\partial M}{\partial\mu}M^{-1}\frac{\partial M}{\partial\mu}M^{-1}\frac{\partial M}{\partial\mu}\right).
\end{eqnarray} 
Five matrix inversions per random vector are necessary here, while for the 6th
order, 12 matrix inversions are needed. Depending on quark mass, temperature
and particular operator, different numbers of random vectors are needed to
obtain that the errors arising from the stochastic estimator are smaller than
or of the same magnitude as the statistical fluctuations within the ensemble.

\section{Pressure and densities}
In this section, we will first show  results for the coefficients, then
use them in computing pressure and quark number densities. 

\begin{figure}
  \centering
  \includegraphics[width=0.48\linewidth]{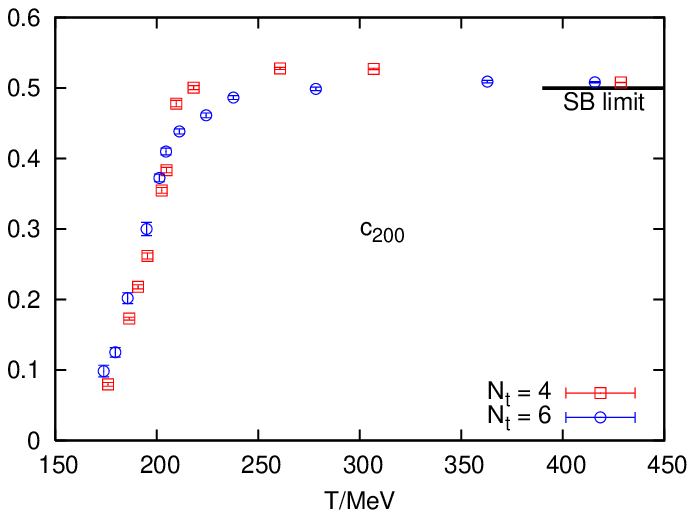}
  \includegraphics[width=0.48\linewidth]{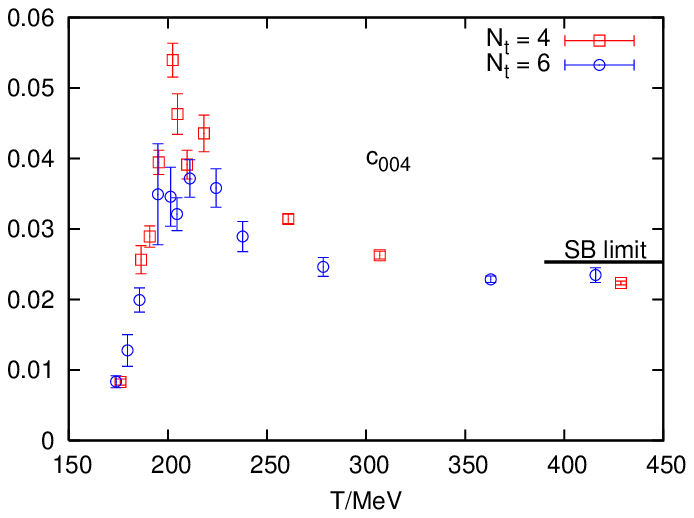}
  \caption{\(c_{200}\) on the left and \(c_{004}\)on the right for \(N_\tau=4\)
    and \(6\). The second order coefficients increase rapidly from confined
    phase to deconfined phase at around \(200\) MeV, while the fourth order
    ones develop a peak there. Stephen-Boltzmann limits of the free case for
    the action that we use are marked for both quantities and matched very well
    in the high temperature region. }
  \label{fig1}
\end{figure}

\begin{figure}
  \begin{minipage}[t]{0.48\linewidth}
        \centering
        \includegraphics[width=0.99\linewidth]{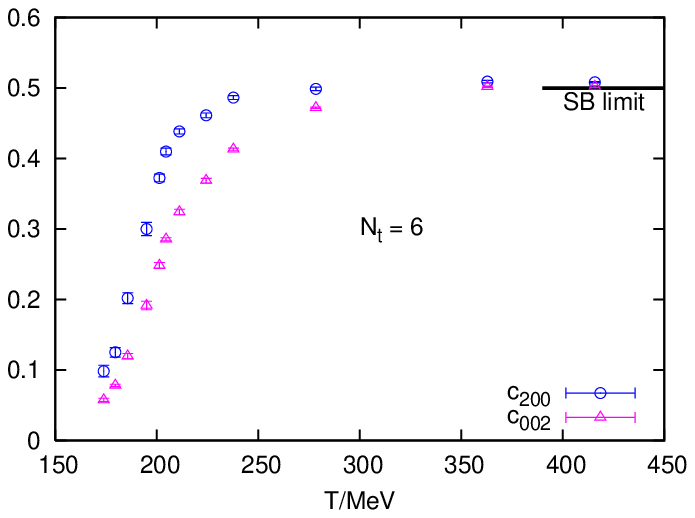}  
        \caption{Second order coefficients \(c_{200}\) and \(c_{002}\) for
          \(u\) and \(s\) quark respectively on \(N_{\tau}=6\) lattice.}
        \label{fig2}
  \end{minipage}
  \begin{minipage}[t]{0.48\linewidth}
        \centering
        \includegraphics[width=0.99\linewidth]{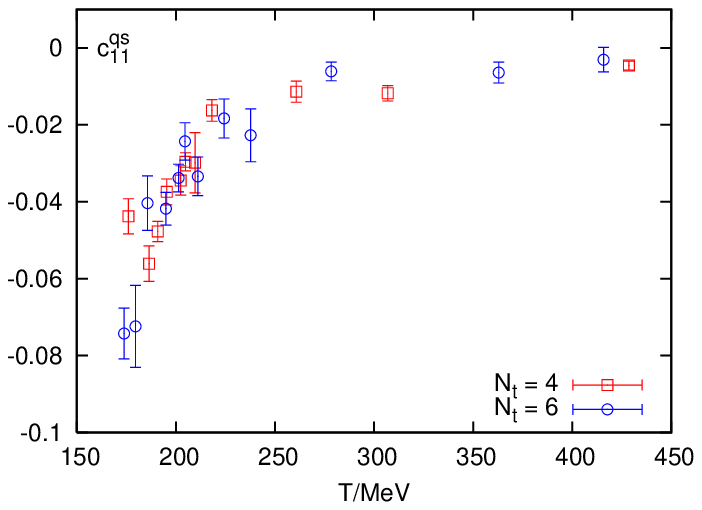}  
        \caption{\(c^{qs}_{11}\) on \(N_{\tau}=4\) and \(6\) lattices. }
        \label{fig3}
  \end{minipage}
\end{figure}

In Fig. \ref{fig1}, 
we show the coefficients \(c_{200}\) and \(c_{004}\) on both \(N_{\tau}=4\) and
\(6\) lattices. \(c_{200}\), also known as the fluctuation in \(u\) (\(d\))
quark number density, increase rapidly through the phase transition region. As
one can see, the lattice cut-off effect is small and seems to be under control.
Results for $c_{002}$ from $N_\tau=8$ lattices \cite{HotQCD} further support
this statement. The fourth order coefficient \(c_{4}\) shows a pronounced peak
around $T_c$. To compare the quark mass dependence, \(c_{200}\) and \(c_{002}\)
for \(u\) and \(s\) quarks respectively are shown in Fig. \ref{fig2}. The slope
is steeper for light than for the strange quarks, which indicates a stronger
sensitivity to the chiral transition for lighter quark masses.
We also show \(c^{qs}_{11}\) in Fig. \ref{fig3}, which approaches zero from
below in the high temperature limit.

\begin{figure}
\includegraphics[width=0.49\textwidth]{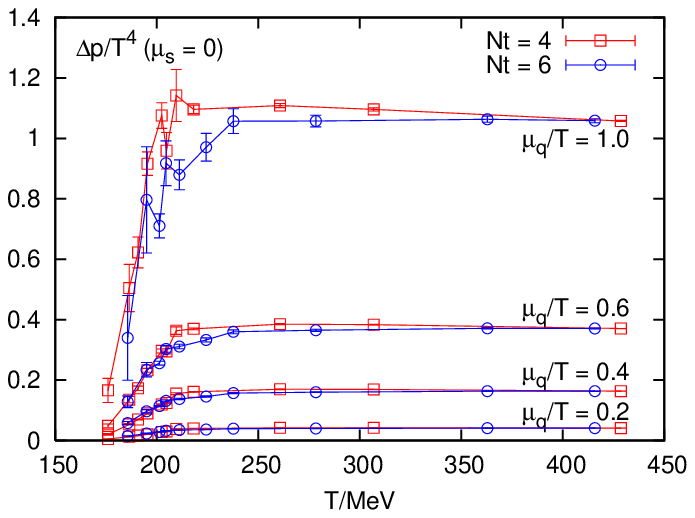}
\includegraphics[width=0.49\textwidth]{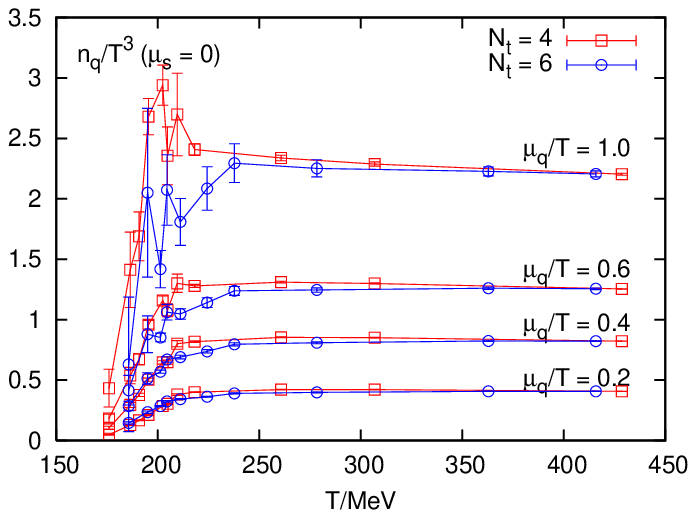}
\caption{Pressure \(\Delta p/T^{4}\) and light quark number density
  \(n_{q}/T^{3}\) at \(\mu_{s}=0\) and \(\mu_{q}/T=0.2\) , \(0.4\), \(0.6\) and
  \(1.0\). Small differences are observed between \(N_{\tau}=4\) and \(6\),
  especially when \(\mu_{q}/T\) is small. Light quark number density seems to
  develop a peak around \(200 \)MeV when \(\mu_{q}/T\) increases.}
\label{fig4}
\end{figure}

Combining all the measured coefficients, we obtain pressure and number density
according to formula (\ref{eq:PTaylor}) and (\ref{eq:ns}).  In Fig. \ref{fig4},
we show the pressure difference \(\Delta p/T^{4}\) and light quark number
density \(n_{q}/T^{3}\) at finite light quark chemical potential but zero
strange quark chemical potential \(\mu_{s}=0\), up to the 4th order.
This should be compared to the pressure at vanishing chemical potential
\cite{Jan}, which rises rapidly to a value of about \(p/T^4\approx 14\) above
the transition. The finite density contribution to the pressure adds to this
less than 10\% for \(\mu_q/T<1\).

\section{Hadronic fluctuations at zero and non-zero chemical potential}
Fluctuations of charge densities $n_{B,S,Q}$ are related by the fluctuation
dissipation theorem to the second derivatives of the partition function with
respect to the corresponding chemical potentials $\mu_{B,S,Q}$. Here $B,S,Q$
denote baryon number, strangeness and electrical charge, respectively. Using
Eqs. (\ref{eq:chempot}) we can rearrange the expansion coefficients
$c^{uds}_{ijk}$ of the pressure to get the coefficients of an expansion in
$\mu_{B,S,Q}$, defined as
\begin{equation}
\frac{p}{T^{4}} = 
\sum_{i,j,k}c^{BSQ}_{ijk}(T)
\left(\frac{\mu_{B}}{T}\right)^{i}
\left(\frac{\mu_{S}}{T}\right)^{j}
\left(\frac{\mu_{Q}}{T}\right)^{k}. 
\label{eq:PTaylorBSQ}
\end{equation}
E.g., the following two relations hold for $c^{BSQ}_{200}\equiv c^{B}_2$ and
$c^{BSQ}_{400}\equiv c^B_4$
\begin{equation}
c^{B}_2
=\frac{1}{9}\left(c^{qs}_{20}
+c^{qs}_{11}+c^{qs}_{02}\right), \qquad 
c^{B}_4
=\frac{1}{81}\left(c^{qs}_{40}
+c^{qs}_{31}+c^{qs}_{22}+c^{qs}_{13}+c^{qs}_{04}\right).
\end{equation} 

\begin{figure}
\includegraphics[width=0.49\textwidth]{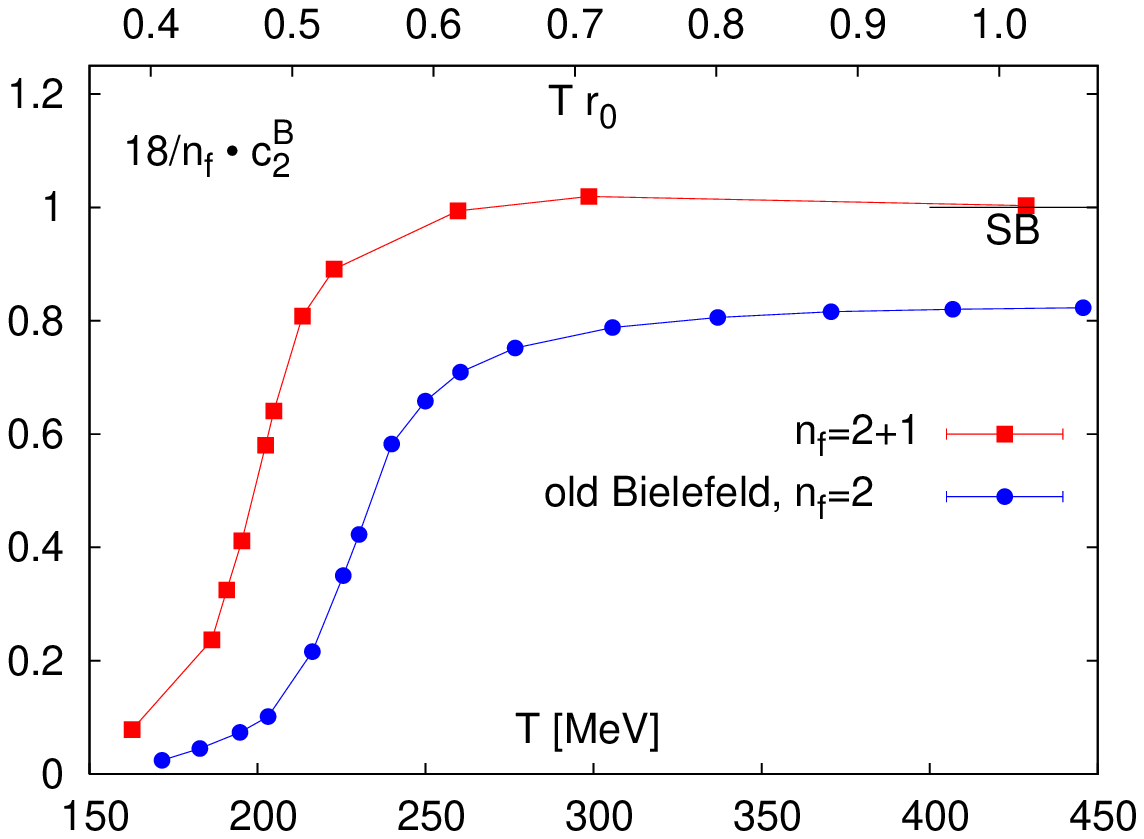}
\includegraphics[width=0.49\textwidth]{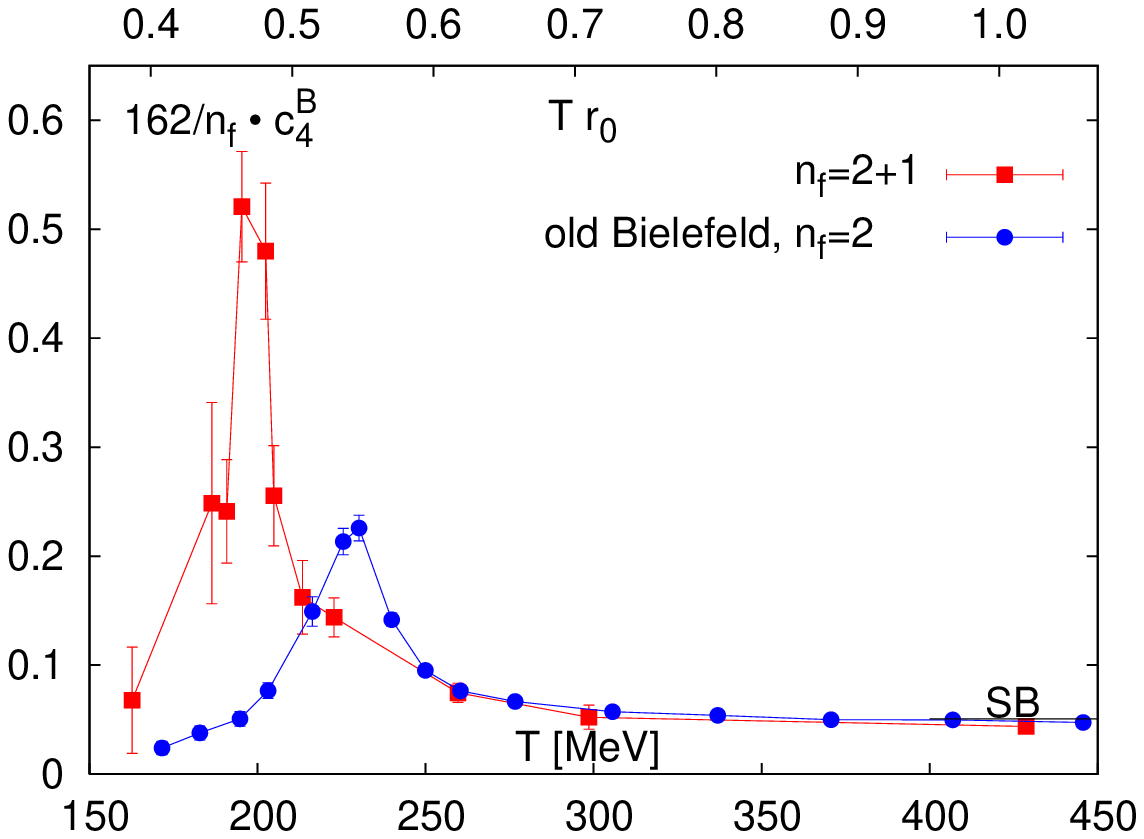}
\caption{Quadratic and quartic baryon number fluctuations at vanishing net
  density as function of temperature. Preliminary data from (2+1)-flavor
  simulations with almost realistic quark masses are compared with previous
  2-flavor simulations [2]. Both results have been obtained on $16^3\times 4$
  lattices. }
\label{fig:cB}
\end{figure}
\begin{figure}
\includegraphics[width=0.49\textwidth]{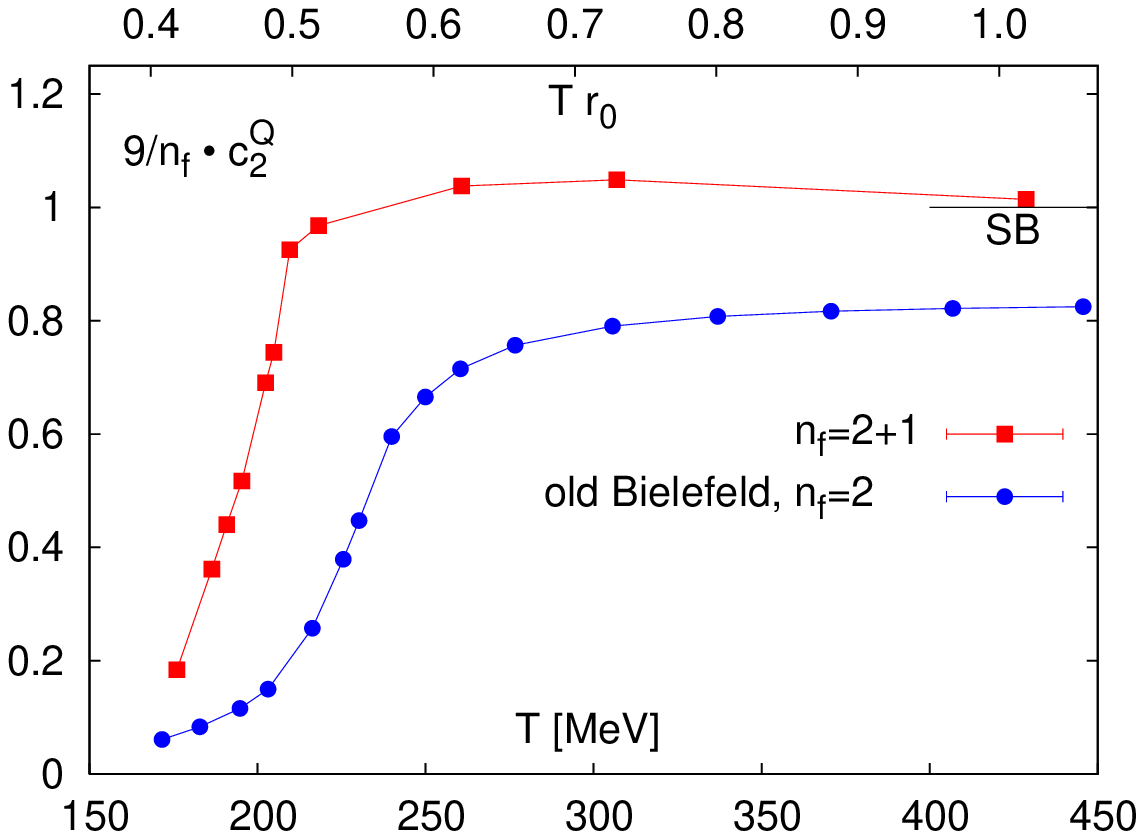}
\includegraphics[width=0.49\textwidth]{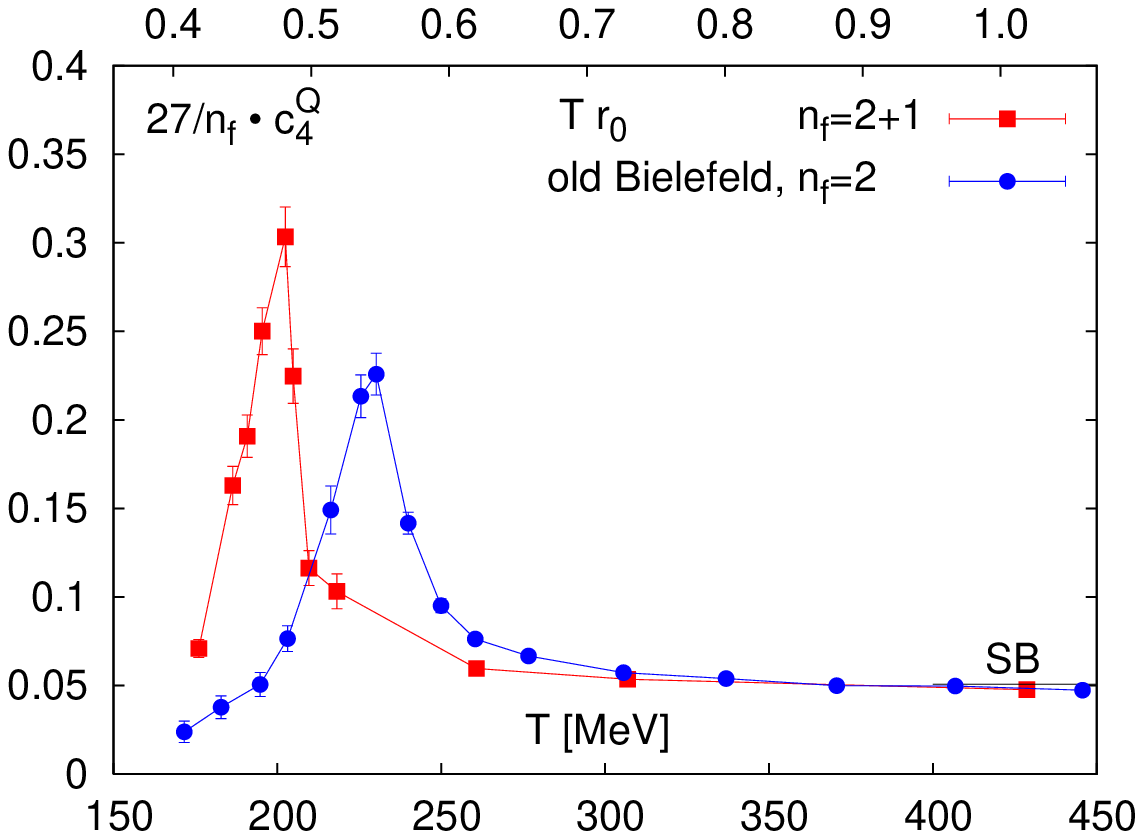}
\caption{Quadratic and quartic electric fluctuations at vanishing net density
  as function of temperature. Preliminary data from (2+1)-flavor simulations
  with almost realistic quark masses are compared with previous 2-flavor
  simulations [2]. Both results have been obtained on $16^3\times 4$ lattices.}
\label{fig:cQ}
\end{figure}
In Fig.~\ref{fig:cB} we show the first two diagonal expansion coefficients in
$\mu_B/T$ as function of temperature, which can also be interpreted as the quadratic
and quartic baryon number fluctuations. We compare our preliminary results for
(2+1)-flavor and almost realistic quark masses to earlier results with 2-flavor
and a pion mass $m_\pi\approx 700 MeV$ \cite{Allton:2005gk}. The normalization
is such that in both cases the same Stefan-Boltzmann value for large temperatures
is reached, i.e. we have divided by the number of flavors. An obvious shift in
the curves reflects the shift in the transition temperature from about 220~$MeV$
to 200~$MeV$. Moreover the sudden change in the quadratic fluctuations is
more pronounced for the smaller masses and the Stefan-Boltzmann value is
reached faster. Correspondingly, the peak in the quartic fluctuations is higher
for smaller masses.

The expansion coefficients in $\mu_S/T$ are identical to that in $\mu_s/T$ --
although the strangeness chemical potential differs from the strange quark
chemical potential by a different sign -- and are shown in Fig.~\ref{fig1} and
\ref{fig2}. In Fig.~\ref{fig:cQ} we show the first two diagonal expansion
coefficients in $\mu_Q/T$. The qualitative picture is very similar to $\mu_B/T$
although the quark mass dependence of the peak height is significantly weaker.

Using the expansion coefficients in $\mu_{B,S,Q}/T$, one can construct hadronic
fluctuations at non-zero baryon number density. Up to fourth order correction
in $\mu_B/T$ we have the following relations for baryon number, strangeness and 
electric charge fluctuations $\chi_{BSQ}$,
\begin{eqnarray}
  \frac{\chi_B(\mu_B/T)}{T^2} &=& 2c^B_2 
  + 12c^B_4\left(\frac{\mu_B}{T}\right)^2 
  + \mathcal{O}\left[\left(\frac{\mu_B}{T}\right)^4\right]\\
  \frac{\chi_S(\mu_B/T)}{T^2} &=& 2c^S_2 
  + 2c^{BS}_{22}\left(\frac{\mu_B}{T}\right)^2 
  + \mathcal{O}\left[\left(\frac{\mu_B}{T}\right)^4\right]\\
  \frac{\chi_Q(\mu_B/T)}{T^2} &=& 2c^Q_2 
  + 2c^{BQ}_{22}\left(\frac{\mu_B}{T}\right)^2 
  + \mathcal{O}\left[\left(\frac{\mu_B}{T}\right)^4\right] .
\end{eqnarray}

\begin{figure}
\includegraphics[width=0.49\textwidth]{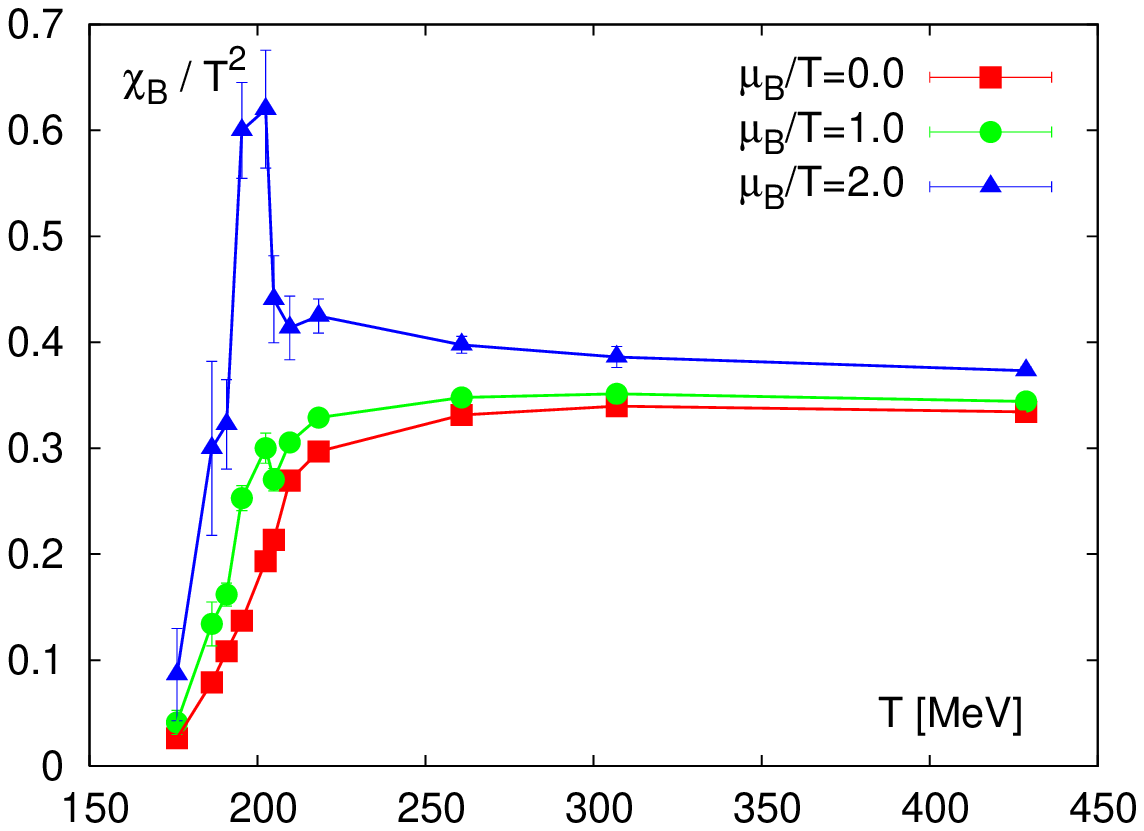}
\includegraphics[width=0.49\textwidth]{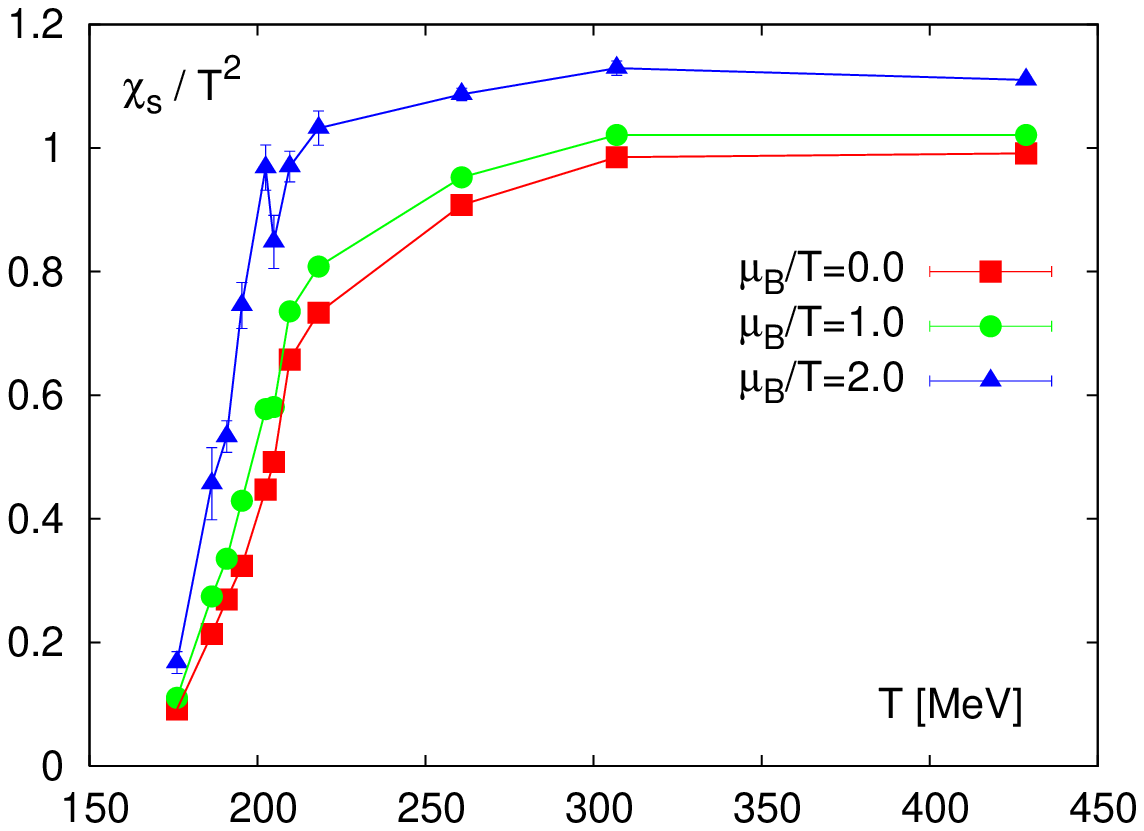}
\caption{Baryon number and strangeness fluctuations at finite baryon number
  density, controlled by a finite baryon chemical potential. Results are
  correct up to fourth order corrections in chemical potential and have been
  obtained on $16^3\times 4$ lattices.}
\label{fig:chi}
\end{figure}

In Fig.~\ref{fig:chi} we show baryon number and strangeness fluctuations at
finite baryon number density. It is obvious that both quantities are developing
a peak for increasing $\mu_B/T$. However, the peak in $\chi_B$ is much more
pronounced since this quantity eventually diverges at the critical point in the
$(T-\mu_B)$-plane. As we anticipated from Fig.~\ref{fig:cB}, the peak height in
$\chi_B$ is about twice as large as in earlier calculations with larger quark
masses \cite{Allton:2005gk}. Note that higher order corrections are still
important, especially the position of the peak will be $\mu_B$-dependent only
by including the next higher order. This has to be analyzed in more detail and
eventually will allow to limit the range of values for $\mu_B/T$ where the
leading order result is reliable.

\begin{figure}
\includegraphics[width=0.49\textwidth]{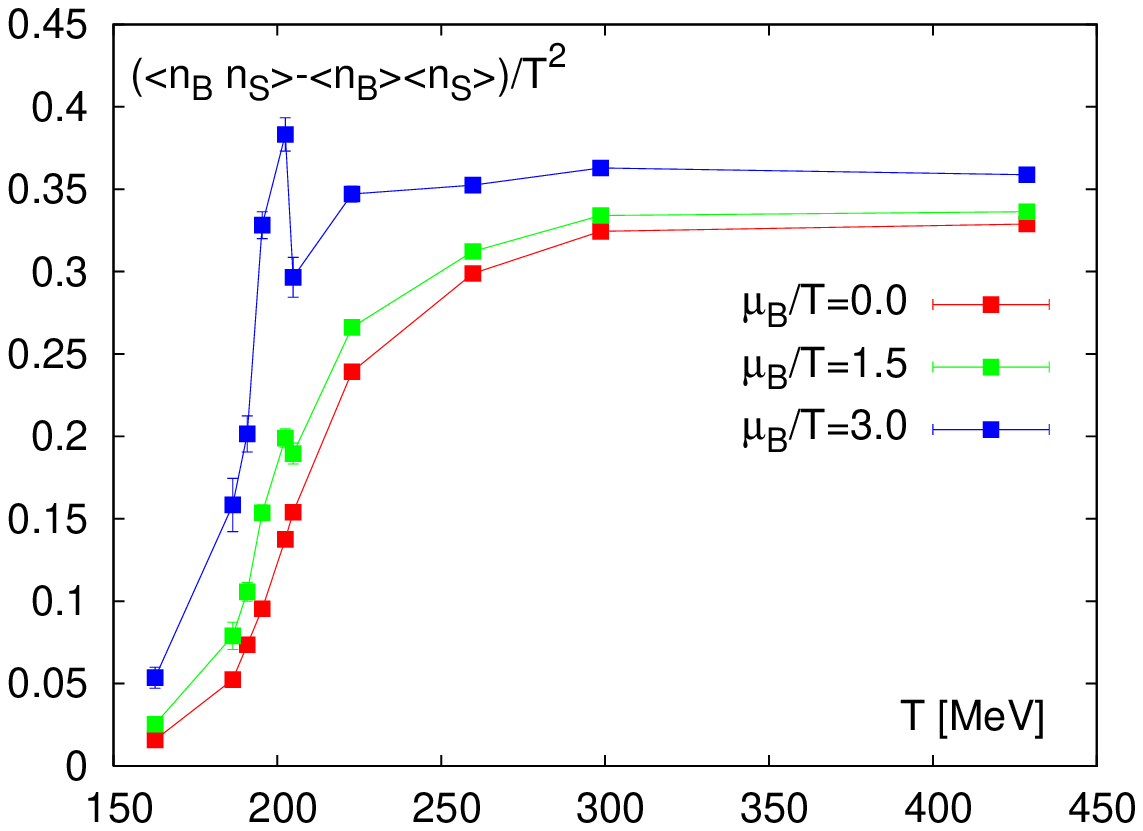}
\includegraphics[width=0.49\textwidth]{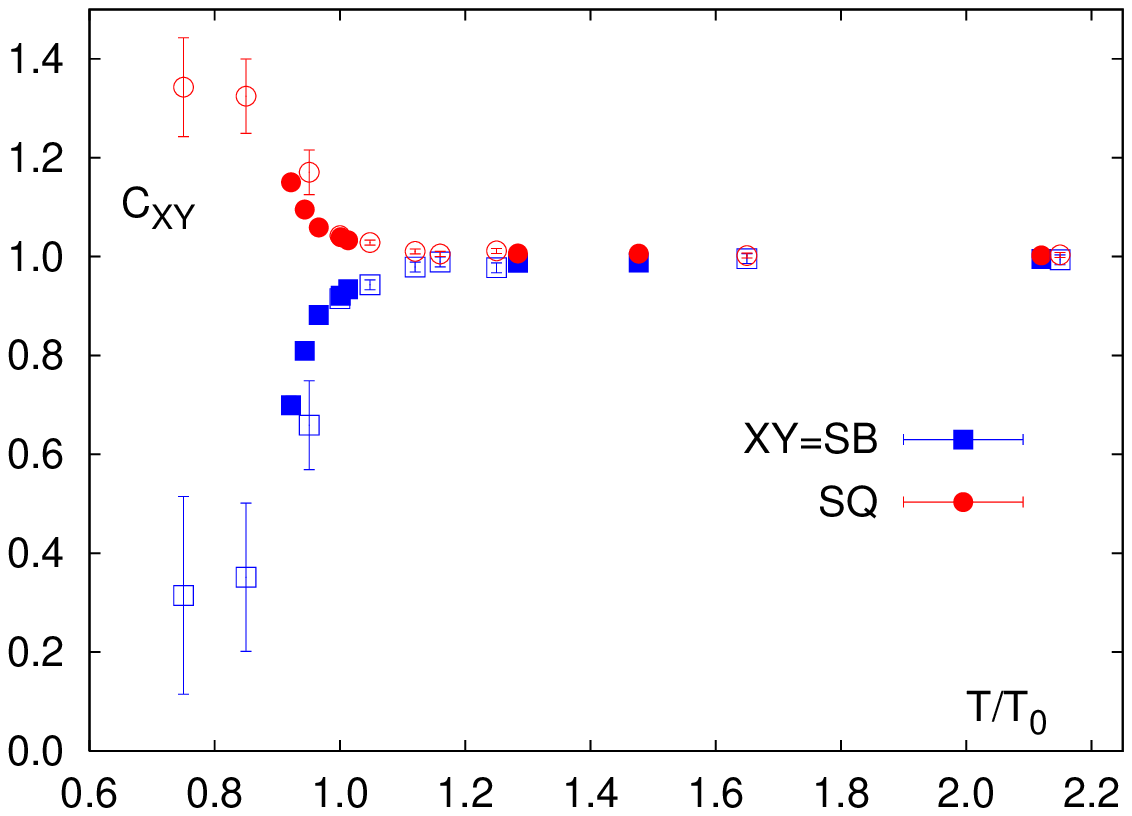}
\caption{Correlation between baryon number and strangeness for several values
  of the baryon chemical potential from $16^3\times 4$ lattices (left) and the
  linkage between baryon number and electric charge with strangeness respectively
  (right). On the right panel we compare or preliminary data (full symbols) to
  previously obtained results from partially quenched calculations (open
  symbols) [7], both obtained on $N_\tau=4$ lattices.}
\label{fig:corr}
\end{figure}

The off-diagonal coefficients in Eq.~\ref{eq:PTaylorBSQ} are usually connected to
correlations between baryon number, strangeness and electrical charge.
The correlation of baryon number and strangeness can be expressed in terms of 
expansion coefficients as
\begin{equation}
\frac{1}{T^2}\left(\left< n_B n_S \right>-\left< n_B \right>\left< n_S \right>\right)
= c^{BS}_{11}+3c^{BS}_{31}\left(\frac{\mu_B}{T}\right)^2
+\mathcal{O}\left[\left(\frac{\mu_B}{T}\right)^4\right] 
\end{equation}
and is shown in Fig.~\ref{fig:corr}. 
We find that also this quantity is developing a peak for increasing chemical
potential, thus the enhanced correlations suggest the vicinity of a critical
point. Another interesting quantity is the ``linkage'' of strangeness and
baryon number or electric charge \cite{Gavai:2005yk}, which is defined as
$C_{SX}=c^{SX}_{11}/c^{S}_2$, where $X=B,Q$. It is known to be a robust
quantity, i.e. the cut-off effects are small. In Fig.~\ref{fig:corr} (right) we
compare our preliminary results with almost realistic quark masses with
previously obtained partially quenched results and slightly larger light quark
masses \cite{Gavai:2005yk}. The two calculations show good agreement, thus also
the quenching and quark mass effects seem to be small in this quantity.  Both
results on correlation and linkage between the different quantum numbers
suggest that the basic charges are carried by quasi-free quark directly above
the transition. This seems to rule out the existence of bound states as
dominant degrees of freedom in this regime \cite{Shuryak:2004tx}.

\section{Conditions at heavy ion colliders and constrained densities}
In general, the pressure, or higher derivatives of the partition functions
with respect to chemical potentials, are dependent on at least 3 variables
$\mu_{u,d,s}$ or equivalently $\mu_{B,S,Q}$. So far we chose $\mu_B>0$, while
holding $\mu_S=\mu_Q=0$.  To compare with experiment, for instance heavy ion
collisions, the chemical potentials might need to be adjusted to meet the
conditions of particular event-by-event fluctuation analyzes
\cite{Begun:2006uu}. A very natural choice of the chemical potentials is
to constrain the strange quark density to zero. Due to the existence of non
zero off-diagonal coefficients in Eq.~\ref{eq:PTaylorBSQ} we find an increasing
strangeness with increasing $\mu_B$, even for $\mu_S=0$. In heavy ion experiments
the total strangeness is zero. Below we outline a procedure to constrain
the net strange quark number density $n_s$ to zero, subsequently order by order 
in our $\mu_B$ expansion. The procedure can be easily generalized to constrain 
other charge densities to arbitrary values. This might be of importance, since
experiments are often restricted to certain rapidity windows, which may alter 
expectation values of charge densities.

\begin{figure}
\centering
\includegraphics[width=0.49\textwidth]{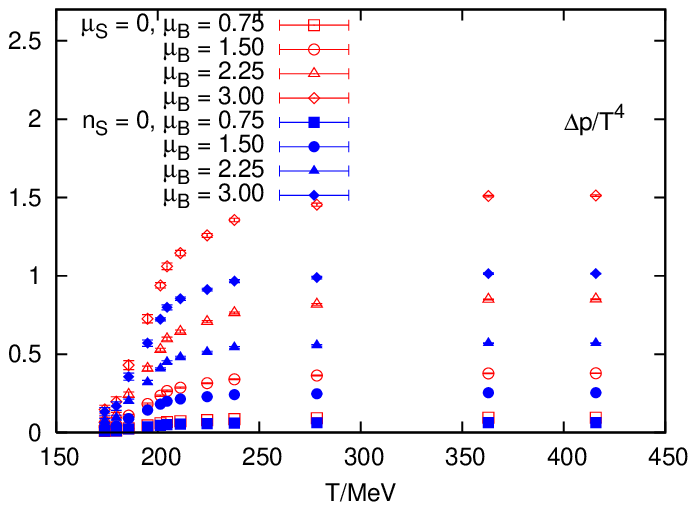}
\includegraphics[width=0.49\textwidth]{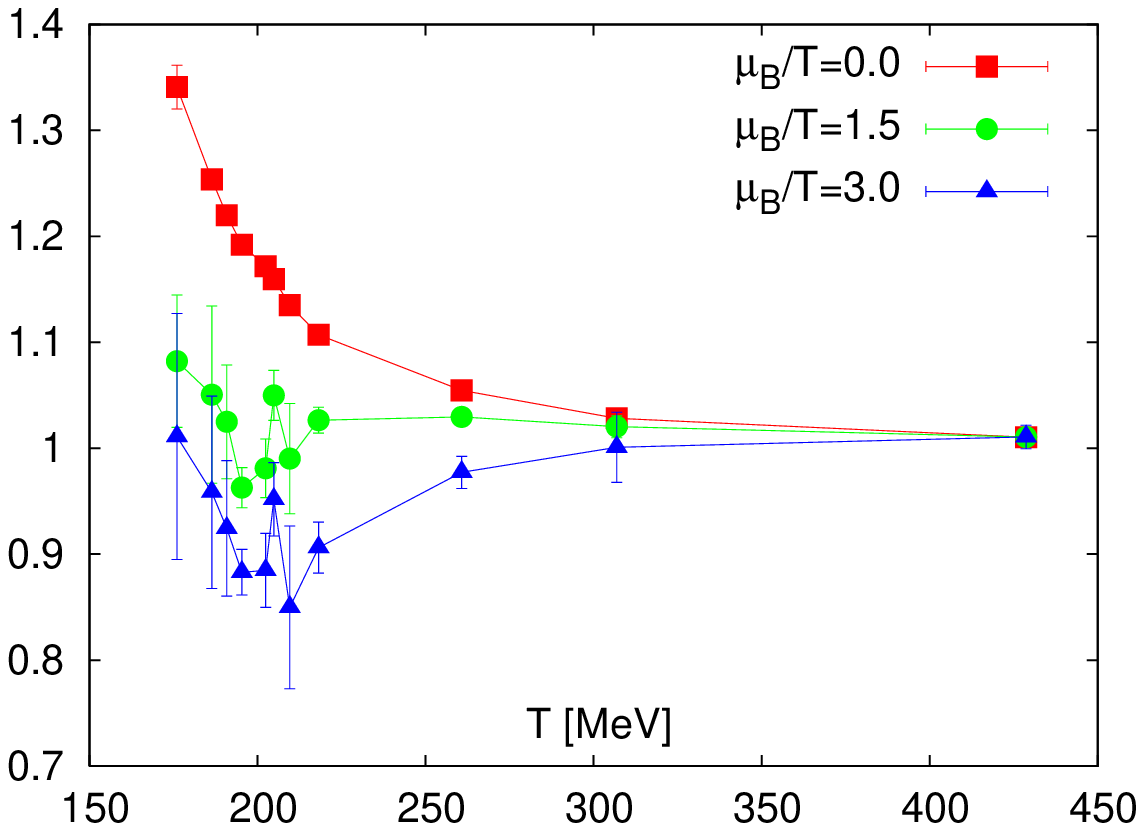}
\caption{The pressure \(\Delta p/T^{4} \) up to the second order for both constraints as labeled
  (left) and the ratio $\mathcal{N} \hat\chi_B/\chi_B$ as explained in the text
  (right) for various values of $\mu_B/T$. The differences between the two 
  constraints are of the order of 30\% for both quantities. Results have been 
  obtained on $24^3\times 6$ lattices (left) and $16^3\times 4$ lattices (right).}
\label{fig:constrain}
\end{figure}

We can express the strange quark number density ($n_s$) in terms of the expansion 
coefficients of the pressure. Up to the 4th order, it reads
\begin{equation}
n_{s}=-n_{S}\left(\hat{\mu}_{B},\hat{\mu}_{S}\right)
= -c^{BS}_{11}\hat{\mu}_{B}    
-2c^{BS}_{02}\hat{\mu}_{S}
- c^{BS}_{31}\hat{\mu}_{B}^{3}                 
-2c^{BS}_{22}\hat{\mu}_{B}^{2}\hat{\mu}_{S}
-3c^{BS}_{13}\hat{\mu}_{B}    \hat{\mu}_{S}^{2}
-4c^{BS}_{04}\hat{\mu}_{S}^{3}\equiv0 ,
\end{equation}
where \(\hat \mu=\mu/T\), which means that the strangeness chemical potential 
\(\mu_{S}\) is no longer a free parameter but depends on
\(\mu_{B}\),
\begin{equation}
\hat{\mu}_{S}\left(\hat{\mu}_{B}\right)
=\left(-\frac{c^{BS}_{11}}{2c^{BS}_{02}}\right)\hat{\mu}_{B}
+\left(\frac{2c^{BS}_{04}{c^{BS}_{11}}^{3}
-3c^{BS}_{02}{c^{BS}_{11}}^{2}c^{BS}_{13}
+4{c^{BS}_{02}}^{2}c^{BS}_{11}c^{BS}_{22}
-4{c^{BS}_{02}}^{3}c^{BS}_{31}}{8{c^{BS}_{02}}^{4}}\right)\hat{\mu}_{B}^{3}
+\mathcal{O}\left(\hat{\mu}_{B}^{5}\right)
.\end{equation} Therefore,  the formula for the pressure is modified to
\begin{equation}
\frac{\Delta p}{T^{4}}
=\left(c^{BS}_{20}-{\frac{{c^{BS}_{11}}^{2}}{4c^{BS}_{02}}}\right)\hat{\mu}_{B}^{2}
+\left(c^{BS}_{40}+{\frac{c^{BS}_{04}{c^{BS}_{11}}^{4 }}{16{c^{BS}_{02}}^{4}}
-\frac{{c^{BS}_{11}}^{3}c^{BS}_{13}}{8{c^{BS}_{02}}^{3}}+\frac{{c^{BS}_{11}}^{2}c^{BS}_{22}}{4{c^{BS}_{02}}^{2}}
-\frac{c^{BS}_{11}c^{BS}_{31}}{2c^{BS}_{02}}}\right)\hat{\mu}_{B}^{4}
+\mathcal{O}\left(\hat{\mu}_{B}^{6}\right),
\end{equation}
which contains off-diagonal coefficients \(c_{11}\),\(c_{13}\), etc. On the
quark level those coefficients are generally small numbers since they are not
present in the free theory. However, on the hadronic level they contain the
diagonal strange quark coefficients which have -- at least in leading order --
a non-zero Stefan-Boltzmann limit. Hence the constraints \(n_{S}=0 \) and
\(\mu_{S}=0\) lead to a quite different dependence of the pressure on $\mu_B/T$,
as can be seen in Fig.\ref{fig:constrain} (left).
The difference is almost negligible, when performing an expansion in the light
quark chemical potential $\mu_q/T$ instead.  It is interesting to mention that
with the constraint $n_s=0$, the pressure expansion in $\mu_q/T$ and $\mu_B/T$
are identical up to a trivial factor between the two chemical potentials, i.e.
the relation $\mu_B=3\mu_q$ holds in this case and we have
\begin{equation}
\left. \Delta p/T^4(\mu_q/T)\right|_{n_s=0} 
\equiv 
\left. \Delta p/T^4(\mu_B/T,\mu_Q=0)\right|_{n_S=0}.
\end{equation} 

We have also computed the constrained baryon number fluctuations at finite
baryon chemical potential $\hat\chi_B$.  Qualitatively, the two cases of
$\mu_S=0$ and $n_S=0$ are very similar. However, it is interesting to remark
that the two cases reach different Stefan-Boltzmann limits for high
temperatures ($T\to\infty$). Taking this into account we show in
Fig.~\ref{fig:constrain} (right) the ratio $\mathcal{N}\hat\chi_B /\chi_B$,
where $\mathcal{N}$ is the ratio of the corresponding Stefan-Boltzmann values.
As one can see, the difference below $T_c$ is as high as 30\%.

\section{Summary and conclusions}
We have presented a method to rigorously compute corrections to bulk
thermodynamic quantities at non vanishing chemical potential, by performing a
Taylor expansion in $\mu/T$. Our new preliminary results improved previous
calculations in many ways: we went to smaller quark masses, finer lattice
spacings and 2+1 dynamical quark flavor. We also showed how to calculate
various hadronic fluctuations, starting from a theory which naturally is
formulated in terms of quark fields, as QCD is. The Taylor expansion method
provides a variety of input to heavy ion phenomenology.

Our findings are that the finite chemical potential contribution to the pressure
is blow 10\%, up to a chemical potential of $\mu_B/T<3$ and that various
hadronic fluctuations develop a peak with increasing baryon chemical potential.
This seems to hold true also for strangeness fluctuations, although the
peak is much less pronounced in this case. Correlations between strangeness and
other charges increase as well when approaching the critical point.

\section*{Acknowledgments}
We would like to thank all members of the RBC-Bielefeld Collaboration for
helpful discussions and comments. The work has been supported in parts by the
U.S. Department of Energy under Contract No. DE-AC02-98CH10886 and by the
Deutsche Forschungsgemeinschaft under grant GRK 881. Numerical simulations have
been performed on the QCDOC computer of the RIKEN-BNL research center the DOE
funded QCDOC at BNL and the APEnext at Bielefeld University.

\end{document}